\documentstyle[preprint,aps,psfig]{revtex}

\begin{document}
\newcommand{\beq}{\begin{equation}}
\newcommand{\eeq}{\end{equation}}
\newcommand{\bea}{\begin{eqnarray}}
\newcommand{\eea}{\end{eqnarray}}
\newcommand{\rv}{\rangle}
\newcommand{\lv}{\langle}

\draft
\tighten
\preprint{IC/96/105}

\title{ON SUPERSYMMETRY AT HIGH TEMPERATURE}

\author{ Borut Bajc \thanks{ borut.bajc@ijs.si}}
\address{International Center for Theoretical Physics,
34100 Trieste, Italy, {\rm and} \newline J. Stefan Institute, 1001 
Ljubljana, Slovenia   } 
\author{ Alejandra Melfo \thanks{
melfo@stardust.sissa.it}}
\address{International School for Advanced Studies,  34014 Trieste, Italy,
 {\rm and} \newline CAT, Universidad de Los
Andes, M\'erida 5101-A, Venezuela}
\author{ Goran Senjanovi\'c \thanks{ goran@ictp.trieste.it}}
\address{International Center for Theoretical Physics,
34100 Trieste, Italy }

\maketitle

\begin{abstract}

While it is possible to find examples of field theories with a spontaneously
 broken symmetry  at high temperature, in renormalizable supersymmetric
 models any internal symmetry gets always restored. Recently, a
 counterexample was suggested in the context of nonrenormalizable
 supersymmetric theories. We show that non negligible higher
 loop effects actually restore the symmetry, without compromising the
 validity of perturbation theory.  We give some arguments as to why the 
proposed mechanism  should not work in general.

\end{abstract}

\section{Introduction}
\label{intr}

In general, whether or not symmetries in field theory are broken at 
high temperature, is a 
dynamical question which depends on the parameter space of the theory in 
question. In spite of one's intuitive expectations on symmetry 
restoration \cite{kl72}, based on daily experience and proven correct
 in the simplest field theory systems \cite{kl72,w74,dj74}, one can easily 
find examples with 
symmetries  broken at high temperature \cite{w74,ms79}. This is an 
important issue, due to its possible role in the production of 
topological defects in the early universe. Symmetry nonrestoration at 
high temperature may provide a way out of both the domain wall 
\cite{ds95,dms96} and the monopole problem \cite{lp80,sss85,dms95}.

A simple example of broken symmetry at high temperature is provided by 
supersymmetry. Due to the different boundary condition for bosons and 
fermions in thermal field theory, supersymmetry is automatically broken 
at any non-zero temperature. However, for the issue of topological 
defects, one would like to know what happens to internal symmetries 
in the context of supersymmetries. This question is nontrivial due to the 
highly constrained structure of SUSY models. It has been addressed 
carefully more than ten years ago \cite{m84}: in contrast 
with the non-supersymmetric case, it was shown that supersymmetry 
necessarily implies restoration of  internal symmetries at high temperature.
At least, this is what happens in renormalizable theories.

Recently, this conclusion was questioned by Dvali and Tamvakis 
\cite{dt96} precisely by resorting to non-renormalizable potentials. They 
present an explicit example in which the inclusion of a quartic term in 
the superpotential allows apparently for  non vanishing vevs at high 
temperature. Stimulated by their interesting suggestion, we have analyzed 
carefully their example, arriving however to the opposite conclusion. 
What happens, and what will be explained in detail below, is that the 
one-loop approximation used by them becomes invalid precisely due to the 
non-renormalizable nature of the superpotential. We find two-loop effects 
actually dominating the one-loop ones, and leading to symmetry 
restoration. We must stress that this is not due to a breakdown of 
perturbation theory, but rather a general feature of theories with more 
than one coupling, and in this sense it is completely analogous to the 
well known Coleman-Weinberg idea \cite{cw73}. There, the one loop 
contribution to the Higgs  self coupling due to gauge interactions may be as 
large as (if not larger than) the tree-level one. We believe this point 
will become clearer after the detailed discussion of our results.

The bulk of this paper is devoted precisely to this, in our opinion, 
important question. In the next section we give first a brief review of 
Dvali and Tamvakis work, and then present our findings. The idea raised 
in \cite{dt96} can in fact be shown to provide a possible new and general 
mechanism, completely independent of SUSY, for symmetry nonrestoration at 
high temperature. We show in section \ref{ci}, on equally general grounds, 
why it cannot work, due to the necessarily dominating higher-loop effects.  

\section{The example: non-renormalizable Wess-Zumino model}
\label{bi}

We take here the prototype model for symmetry nonrestoration of Ref. 
\cite{dt96}, which is basically a Wess-Zumino model with a discrete 
symmetry ${\bf D}: \Phi \to -\Phi$ and the addition of a non-renormalizable 
interaction term:

\beq
W = -{1 \over 2} \mu \Phi^2 + {1 \over 4 M} \Phi^4\;,
\label{spot}
\eeq

\noindent
where $M\gg\mu$. This leads to the scalar potential

\bea
V &=& |\phi|^2 |-\mu + {\phi^2 \over M}|^2 \nonumber \\
 &=& {\mu^2 \over 2} (\phi_1^2 + \phi_2^2)  - 
{\mu \over 2M} (\phi_1^4 - \phi_2^4) + 
{1 \over 8 M^2} (\phi_1^2 + \phi_2^2)^3\;, 
\label{pot} 
\eea

\noindent
where $\phi = (\phi_1 + i \phi_2)/\sqrt{2}$ is the scalar component of the 
chiral Wess-Zumino superfield $\Phi$. Notice that $\phi_1$ has a negative
 quartic self coupling. 
At $T=0$, as usual, one finds a set of two degenerate minima: 
$\lv\phi\rv = 0$ and $\lv \phi \rv^2 = \mu M$. To see what happens at 
high T, in Ref. \cite{dt96} the usual 1-loop induced correction to the
 effective potential is computed

\beq 
\Delta V_{\rm 1-loop}(T) = {T^2 \over 8} |
{\partial^2 W \over \partial \phi^2}|^2 =
 {T^2 \over 8} |-\mu + {3 \phi^2 \over M} |^2
\label{Tpot1}
\eeq

\noindent or

\beq
\Delta V_{\rm 1-loop}(T) ={T^2 \over 8} \left[\mu^2 - 3{\mu\over M} 
(\phi_1^2 - \phi_2^2) + {9 \over 4 M^2} (\phi_1^2 + \phi_2^2)^2 \right]\;.
\label{Tpot2}
\eeq

\noindent
Dvali and Tamvakis conclude  
that for $ M^2 \gg T^2 \gg \mu M$, one gets $\lv \phi \rv^2 \neq 0$, as is
 immediately clear from (\ref{Tpot1}).  

Before we move on to question this statement, let us see what really is going
 on up to this point. As is transparent from (\ref{Tpot2}), one can
 attribute (as usual) the symmetry breaking to the negative $T^2$ mass term
 for $\phi_1$. However, the quartic self coupling of $\phi_1$ in (\ref{pot})
 being negative, one cannot ensure a nonvanishing vev. It is necessary to
 assume that either the $\phi_1^6$ term in (\ref{pot}) or  the  $\phi_1^4$
 term in (\ref{Tpot2}) (both positive, but suppressed by $M^2$), dominates 
over the $(\mu/M) \phi_1^4$ term, in order for $\phi_1$ to have a vev. 

Now, as we have required $T^2 \gg \mu M$, this is perfectly acceptable. 
But this is precisely where the problem lies: one has assumed that the 
non-renormalizable terms are not small in comparison with the renormalizable
 ones. Notice that this does not put in question the validity of perturbation 
theory, since the $\phi^4$ terms are suppressed by the small parameter 
$\mu/M$. This is the analogy with the Coleman-Weinberg case that we drew
 before. Perturbation theory is perfectly safe, since the next term in the
 series would be of order $\phi^8/M^4$, or $T^2 \phi^6 / M^4$, which are
 strongly suppressed by $T/M \ll 1$ or $\phi/ M \ll 1$.

The question is: what about a term such as $T^4 \phi^2 /M^2$, which is
 obviously much bigger that $(\mu/M) T^2 \phi^2$ ? Notice that this is the only
 relevant term that one could have missed, and whose sign would decide the
 pattern of symmetry breaking, if present.  Once again, the idea is to write 
the expansion in $1/M$, but due to the fact that one has two different 
couplings to start with (namely, $(\mu/M) \phi^4$ and $\phi^6 / M^2$),
 one cannot resort to the usual loop expansion, since $T^4 \phi^2/M^2$ does 
not appear at one loop.  
Again, notice the complete parallel with the Coleman-Weinberg analysis. There,
 assuming a small self coupling $\lambda$ for the Higgs field, one finds an 
important $\phi^4$ term proportional to $g^4$ ($g$ being the gauge coupling 
constant) only at one loop, without implying the failure of perturbation 
theory. 

We have found out that such a term, $T^4 \phi^2 / M^2$, does actually 
appear at the two loop level, through the diagrams depicted in Fig. 1.

Using the superpotential (\ref{spot}) and the usual rules for the evaluation
 of Feynman diagrams in thermal field theory \cite{w74,dj74}, it
 is straightforward to calculate this contribution as

\beq
\Delta V_{\rm 2-loops}(T) = {9 T^4\over 32 M^2 } |\phi|^2 = 
{9 T^4\over 64 M^2} (\phi_1^2 + \phi_2^2)\;.
\label{Tpot2l}
\eeq

We wish to stress again that this term, in the range of parameters considered
($ M^2 \gg T^2 \gg \mu M$), dominates over the mass term in (\ref{Tpot2}),
 and therefore must be taken into account. Since it is positive, the 
conclusion is contrary to the one in Ref. \cite{dt96}: the discrete 
symmetry is restored at high temperature.

As we said before this result is valid up to leading order in an expansion 
in $\phi/M$ and $T/M$. As long as we stay far away from $M$, the perturbation 
theory guarantees symmetry restoration. The reader may still feel uneasy about
 the consistency of calculations performed in a non-renormalizable theory. 
For this reason we have also  performed  the calculations in the 
renormalizable 
version of the theory. That is, as suggested in Ref. \cite{dt96}, one can 
consider a renormalizable superpotential

\beq
W = {\mu \over 2} \Phi^2 + { M_X \over 2} X^2 + \lambda X \Phi^2\;,
\label{rspot}
\eeq
 
\noindent which upon integrating out the heavy field $X$, gives 
(\ref{spot}) after
 identifying $M = M_X/2\lambda^2$. Not surprisingly, considering 
all the graphs and 
in the limit $M >>T$, the same correction (\ref{Tpot2l}) for the effective 
potential is obtained.

\section{General discussion}
\label{ci}

We have seen that, unfortunately, the suggestion put forward by the
authors of  Ref. 
\cite{dt96} does not work. Here we should point out that the question 
raised by them has a much more general significance. Namely, if their
 idea were to work, this would pave a way to a new mechanism for avoiding
 symmetry restoration at high temperature, completely independently 
of supersymmetry.

To explain what is going on, let us recall the basics of  symmetry
nonrestoration at high temperature in renormalizable theories. Consider a 
theory with two scalar fields $\phi_1, \phi_2$ and a potential symmetric under
{\bf D:}${\phi_1\to -\phi_1, \phi_2 \to -\phi_2}$ 

\beq
V(\phi_1, \phi_2) = \sum_{i=1}^2\left(- {m_i^2 \over 2} \phi_i^2 +
 {\lambda_i \over 4} \phi_i^4 \right) - {\alpha \over 2} \phi_1^2 \phi_2^2
+ \beta_1\, \phi_1^3 \phi_2 + \beta_2\, \phi_2^3 \phi_1\;.
\label{toy}
\eeq

The self couplings $\lambda_i$ must be positive for the potential to be 
bounded from below, but one can always 
choose $\alpha >0$, $\beta_1, \beta_2 >0$ in (\ref{toy}), and require 
$\lambda_1 \lambda_2 > \alpha^2$. The high temperature corrections are

\begin{equation}
\Delta V_{\rm 1-loop}(T) = {T^2\over 24} 
\left[(3 \lambda_1 - \alpha)\phi_1^2 + (3 \lambda_2 -
\alpha)\phi_2^2 + 6 (\beta_1 + \beta_2) \phi_1 \phi_2 \right]\;.
\end{equation}
 
By asking, e.g. $\alpha > 3 \lambda_1$, one can keep one of the mass terms
negative at any temperature, thus keeping the symmetry broken at any T.
Notice that the signs of the  scalar interactions and the 
corresponding  T-dependent  terms are equal. We have seen in 
section \ref{bi} that this feature persists in non-renormalizable theories.

In theories with a single field, this mechanism of nonrestoration would 
apparently be impossible, since the self-coupling must be positive in order 
to guarantee the boundedness of the potential. Here precisely enters the 
point of Ref. \cite{dt96}: they have a negative quartic interaction for 
the $\phi_1$ field but the theory is rendered stable through the positive 
non-renormalizable $\phi_1^6$ term. 
 
 One could extend this mechanism to an arbitrary non-supersymmetric theory. 
To see why this cannot work in general, let us take the example of a real
 scalar field with a negative and small quartic self coupling, 
and a discrete symmetry {\bf D:}$\phi\to -\phi$

\beq  
V = {\mu^2 \over 2} \phi^2 - \epsilon \phi^4 + {\phi^{2n+4} \over M^{2n}}\;,
\label{scalarpot}
\eeq

\noindent
where we include the first important non-renormalizable term. The power 
$n$ varies from model to model ($n=1$ in the case discussed above). The 
idea of Ref. \cite{dt96} is based on two important points:
$\epsilon >0$ and $ \epsilon <<1$.

Notice that the non-renormalizable term makes the theory stable independently
 of the sign of $\epsilon$.  At one loop level,  one gets for $T<<M$ the
 corrections

\beq
\Delta V_{\rm 1-loop}(T) = {T^2\over 24} \left[ - 12 \epsilon \phi^2 +
{(2n+4) (2n+3) \over M^{2n} }\phi^{2n+2} \right]\;.
\eeq

The idea is then that the temperature-induced non-renormalizable term is to 
combine with the one coming form the negative self-coupling to induce a vev 
when $\Delta V(T)$ starts to dominate, i.e for $T^2 >> \mu^2$. But of course, 
for this to happen one has to assume that the non-renormalizable 
term is not negligible, i.e. $\epsilon$ very small. Here comes the point: 
if the non-renormalizable term is not negligible, one has to take into
 account its contributions to the thermal mass. This means that the 
expansion cannot end at one loop, but has to be pursued up to $n+1$ loops.
At that level, the ``butterfly'' diagrams with $n+1$ loops and 
two external legs of which Fig. 1 (a) 
is the $n=1$ example, will induce the high 
temperature contribution

\beq
\Delta V_{\rm n+1-loops}|_{\rm mass\; term}(T) = 
{1 \over 2} \left({T^2 \over 12}\right)^{n+1} 
{(2n+4)! \over 2^{n+1} (n+1)!} {1 \over M^{2n}} \phi^2\;.
\label{nmass}
\eeq

Any other term in the expansion of the couplings $1/M^{2n}$ and $\epsilon$ 
will be suppressed. Each loop
 in the diagram will provide a positive contribution $T^2/12$, so the sign of 
(\ref{nmass}) is the sign of the coupling. A positive mass term already 
indicates that the symmetry will be restored, however one should look 
at all the temperature-dependent interactions that follow from the 
non-renormalizable terms. The diagrams that give the dominant $1/M^{2n}$
 contribution to the $\phi^{2m}$ interaction terms are again the
 ``butterflies'' with $2m$ external legs, and they are readily calculated
 
\beq
\Delta V(T) = \sum_{m=1}^{n+1} {(2n+4)!\over (2m)!
(n-m+2)! 2^{n-m+2}} \left( 
{T^2 \over 12} \right)^{n-m+2} {\phi^{2m} \over M^{2n}} \;.
\label{bsf}
\eeq

All the terms of the series have a positive sign, not surprisingly, as 
we mentioned before the high-T contributions carry the sign of the
 coupling constant. Symmetry restoration then follows.

We can easily generalize (\ref{bsf}) to get the ``butterfly'' contribution 
to the high temperature effective potential of  an arbitrary $V(\phi)$: 

\beq
V(\phi,T)= \sum_{m=0}^\infty {1 \over m!} \left({T^2 \over 24} \right)^m
\left\{ {d^{2m}V \over d\phi^{2m}}(\phi) -  
{d^{2m}V \over d\phi^{2m}}(\phi=0)     \right\} \; .
\eeq

\section{Conclusions }

According to our results, the idea of Ref. \cite{dt96} of using higher
dimensional effective interactions to provide nonrestoration of internal 
symmetries in a supersymmetric context seems not to work. This, in turn, 
would confirm the general result  \cite{m84} which was proved only for
 renormalizable supersymmetric theories.

 We have also offered arguments of why we believe this to hold
 in general. However, admittedly, we do not have a rigorous proof of symmetry
 restoration in the multifield nonrenormalizable supersymmetric case. 
The purpose of our paper is in a sense twofold: first, since the issue is
 so important, it was crucial to know whether the discussed mechanism for 
nonrestoration \cite{dt96}   
is valid or not; and second, we hope that it may inspire the reader to
 provide either a rigorous proof of the no-go theorem \cite{m84}, 
or a way out.

\section*{Acknowledgements}

We have  benefited from discussions with Hossein Sarmadi and Francesco 
Vissani. We also thank Gia Dvali and Kyriakos Tamvakis for their interest 
and comments. B.B. thanks ICTP for hospitality, and both ICTP and the 
Ministry of Science and Technology of Slovenia for financial support.

\begin{figure}[h]
\centerline{\psfig{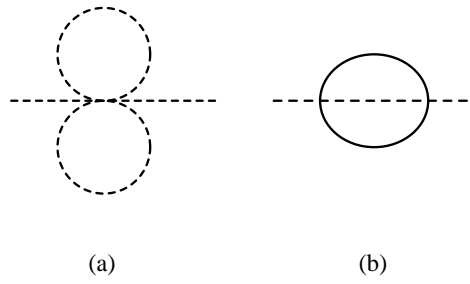}}
\centerline{   }
\centerline{   }
\caption{Feynamn diagrams for the thermal mass correction from the 
non-renormalizable term. Dashed lines represent the scalar boson 
$\phi$, continuos lines represent its fermion counterpart
$\tilde \phi$.}
 \end{figure}

\end{document}